\documentclass[12pt]{article}
\usepackage{amsmath,amssymb,array,cite,latexsym}
\usepackage{graphicx}

\newcommand{\be}{\begin{equation}}
\newcommand{\ee}{\end{equation}}
\def\bea{\begin{eqnarray}}
\def\eea{\end{eqnarray}}

\begin{document}


\begin{center}
\Large{\bf Generalized coherent states for time-dependent and nonlinear Hamiltonians via complex Riccati equations}
\end{center}

\vskip2ex
\begin{center}
Octavio Casta\~{n}os\\
{\footnotesize Instituto de Ciencias Nucleares, Universidad National Aut\'{o}noma de M\'{e}xico,\\ 
Apdo. Postal 70-543, 04510 M\'{e}xico D.F.\\
ocasta@nucleares.unam.mx}\\[2ex]
Dieter Schuch\\
{\footnotesize Institut f\"ur Theoretische Physik, J.W. Goethe-Universit\"at Frankfurt am Main,\\ 
Max-von-Laue-Str. 1, D-60438 Frankfurt am Main, Germany\\
Schuch@em.uni-frankfurt.de}\\[2ex]
Oscar Rosas-Ortiz\\
{\footnotesize Departamento de F\'{i}sica, Cinvestav,\\ 
Apdo. Postal 14-740, 07000 M\'{e}xico, D.F.\\
orosas@fis.cinvestav.mx}
\end{center}

\begin{center}
\begin{minipage}{14cm}
{\footnotesize {\bf Abstract.}
Based on the Gaussian wave packet solution for the harmonic oscillator and the
corresponding creation and annihilation operators, a generalization is
presented that also applies for wave packets with time-dependent width as they
occur for systems with different initial conditions, time-dependent frequency
or in contact with a dissipative environment. In all these cases the
corresponding coherent states, position and momentum uncertainties and quantum
mechanical energy contributions can be obtained in the same form if the
creation and annihilation operators are expressed in terms of a complex
variable that fulfills a nonlinear Riccati equation which determines the
time-evolution of the wave packet width. The solutions of this Riccati
equation depend on the physical system under consideration and on the
(complex) initial conditions and have close formal similarities with general
superpotentials leading to isospectral potentials in supersymmetric quantum
mechanics. The definition of the generalized creation and annihilation
operator is also in agreement with a factorization of the operator
corresponding to the Ermakov invariant that exists in all cases considered.}
\end{minipage}
\end{center}


\section{Introduction}

Shortly after the first communication on wave mechanics \cite{1} Schr\"odinger
himself already considered the possibility of constructing stable wave packet
(WP)-type solutions of his equation that behave closest to the classical
particles in the case of the harmonic oscillator (HO) \cite{2}. The functions
that fulfill this requirement, the Gaussian WPs, are determined by two
parameters, the maximum and the width, where in his case the maximum followed
the classical trajectory and the width was just a constant. This led him to think that the WP could just describe a distribution of matter instead of a mass point, but this hope was destroyed by the rapidly spreading WP for the free motion.

Generalizations of Schr\"odinger's approach were achieved in the description
of coherent light beams emitted by lasers \cite{3,4,5},
considering what is now known as coherent state (CS).  There
are at least three different formal definitions of these states in the
literature \cite{6}: (1) minimum uncertainty CS, meaning Gaussian WPs that
minimize the Heisenberg uncertainty relation; (2) annihilation operator CS and
(3) displacement operator CS. The definition of the annihilation
operator CSs involves the factorization of the Hamiltonian (of the HO) in terms of creation and annihilation operators, a procedure that
was also applied by Schr\"odinger \cite{7}, but already mentioned in principle
by Dirac \cite{8}. All this works perfectly well for at most quadratic and
conserved (time-independent) Hamiltonians with Gaussian WP solutions. However,
this type of WP solutions also exists if the Hamiltonian is no longer a
constant of motion, but explicitly time-dependent (TD), e.g. like in the case
of a HO with TD frequency $\omega(t)$. In this case, a dynamical
invariant (with the dimensions of an action instead of energy) still exists, as
has been shown in a formal mathematical context by Ermakov~\cite{9} already 45
years before quantum mechanics was developed by Schr\"odinger and
Heisenberg. This invariant can be obtained by eliminating the TD frequency
$\omega(t)$ from the Newton equation of motion  for the WP maximum with the
help of an auxiliary equation that, as will be shown later, determines the
time-evolution of the position uncertainty (or WP width). The important role of this Ermakov invariant, particularly in a quantum mechanical context, has been pointed out in the works of Lewis and Riesenfeld \cite{10} and has been exploited by many other authors. In this context, also generalized CSs for these systems have been discussed \cite{11}.

In this work we will give a further generalization that not only covers Hamiltonians with TD (harmonic) potentials, but also TD Hamiltonians that describe dissipative systems where the energy is no longer a constant of motion and other, particularly nonlinear (NL) extensions of the TDSE, that try to include the effect of a dissipative environment into a quantum mechanical context. 

It is assumed that also for these systems a description in terms of pure
states is possible where these pure states fulfill modified SEs (with possibly
non-Hermitian Hamiltonians) that still have exact Gaussian-type WP
solutions. There are several of these approaches in the literature
\cite{12,13,14,15,16} and we will show how some of the
most accepted ones are interrelated and the meaning of their physical interpretation. For some of these approaches, it can be shown \cite{17} that also in these cases an exact dynamical invariant of Ermakov-type exists. This invariant can be expressed in different variables corresponding to the respective Hamiltonian, but if the proper transformations between these approaches are taken into account, they all lead to the same invariant.

This invariant, as in Schr\"odinger's original approach and in the extension
to include TD frequencies for the HO, can also be factorized and, for the quantized version of the invariant, this leads to generalized creation and annihilation operators that are also applicable for open dissipative quantum systems.

A key point of our approach is that we do not express these generalized
creation and annihilation operators in terms of the (real) auxiliary variable
that fulfills the (NL) Ermakov equation, but in terms of a complex variable
that fulfills a quadratic NL Riccati equation. The imaginary part of this
variable satisfying the Riccati equation is uniquely related to the position
uncertainty  or the real Ermakov variable. The particular form of the real
part can change according to the system of interest (conservative or
dissipative). In our generalized formalism, the complex Riccati equation
exists (at least) for HO systems with constant or TD frequency and with or without linear velocity dependent friction forces (where the damped free motion is covered for $\omega \rightarrow 0$). So, the same form of the generalized creation/annihilation operators in terms of the Riccati variable is applicable for all these systems which is not the case if they are expressed in terms of the real Ermakov variable.

A further advantage of the complex Riccati approach over the real Ermakov one
is that in the Riccati case, due to the quadratic nonlinearity (particularly
in the dissipative case), the possibility of pairs of solutions with different
physical properties (connected with phenomena like Hopf bifurcations)
immediately becomes obvious, which is not at all the case if the corresponding
real NL Ermakov equation is considered. The solutions of the Riccati equation
are quite sensitive to the (complex) initial conditions and can, for even
slightly different ones, lead to a qualitatively very different behaviour of the
quantum system. Due to the Riccati equation, formal similarities with the
supersymmetric approach to quantum mechanics \cite{18,19,20} can
also be established and exploited. 

We introduce our generalized CSs in terms of the complex Riccati variable using
the TDSE for the HO with the possibly TD frequency as starting point and show,
that our approach is consistent with the definitions of displacement operator
CSs and annihilation operator CSs. Since we consider general CSs, with
position-momentum correlations, the width of the Gaussian WP solutions is no
longer constant and these CSs do not fulfill the minimum uncertainty
requirement, but the modified one associated with the Schr\"odinger--Robertson
uncertainty relation \cite{21}. 

The approaches to include dissipation and irreversibility that will be
discussed have one thing in common: they lead to Langevin-type evolution
equations (with linear velocity dependent friction force) for the classical
part of the system, determining the motion of the WP maximum. No stochastic
force appears explicitly, but this aspect might be taken into account by non-Hermitian
imaginary contributions to the Hamiltonian in the NLSE approaches.  Two of
these approaches start already on the quantum level and add some (NL) terms to
incorporate the effect of the dissipative environment (leaving the standard
definitions and interpretations of operators and wave functions
unchanged). Two other ones modify the canonical Lagrange/Hamilton formalisms
on the classical level in a way that allows one to include the above-mentioned
friction force, but after quantization caution must be taken with respect to the physical meaning of operators and wave functions in this canonical context. We will show that all the considered approaches share the same Ermakov invariant and how this can be expressed and factorized in physical terms.

In particular, the physically-relevant Riccati equation, its possible
solutions and relations to supersymmetric quantum mechanics will be discussed
and the quantum uncertainties will be expressed in terms of the complex
Riccati variable.

\section{Gaussian wave packets, complex Riccati equations and generalized coherent states}

\subsection{Gaussian wave packets and Ermakov invariant}

In the following, TDSEs with at most
quadratic Hamiltonian (particularly, the HO with constant or
TD frequency $\omega$, and the free motion, $V=0$ in the limit $\omega
\rightarrow 0$) in one dimension shall be considered,
\begin{equation}    
    i\hbar \frac{\partial}{\partial t} \Psi (x,t) = \left\{ -
    \frac{\hbar^2}{2m} \frac{\partial^2}{\partial x^2}
    + \frac{m}{2} \omega^2 x^2  \right\}\Psi (x,t).
\label{mex1}
\end{equation}
In these cases, Gaussian wave packet solutions can be obtained that can be written in the form
\begin{equation} 
\Psi_{WP} (x,t) = N (t)  \exp \left\{ i \bigg[ y(t) \widetilde
        x^2 + \frac{1}{\hbar} \langle p\rangle  \widetilde x + K (t)\bigg]\right\} 
\label{mex2}
\end{equation}
with $\widetilde x = x - \langle x\rangle  = x - \eta$ where the mean value of position is given
by $\langle x \rangle  = \int^{+\infty}_{-\infty} dx \Psi^*x\Psi  = \eta(t)$, $\langle p \rangle  = m  \dot{\eta}$ and
the purely TD function $K(t)$ and normalization factor $N(t)$ (which will be
specified later). The time-dependent coefficient of the quadratic term in the
exponent is assumed to be complex, $y(t) = y_{R} + i y_{I}$ where the
imaginary part is related to the position uncertainty $\langle \widetilde{x}^2 \rangle  = \langle x^2 \rangle  -
\langle x \rangle ^{2}$ via $y_{I} = \frac{1}{4  \langle \widetilde{x}^2 \rangle }$. The maximum of the WP is located at $x = \langle x \rangle (t) = \eta$ and, thus, follows the classical trajectory determined by the Newtonian equation of motion that can be obtained by inserting (2) into (1),
\begin{equation}
\ddot{\eta} + \omega^2(t){\eta}   = 0.
\label{mex3}
\end{equation}
The time-evolution of the WP width can be determined in the same way and is governed by the complex nonlinear Riccati equation
\begin{equation}  
   \left( \frac{2\hbar}{m} \dot y \right) + \left(\frac{2\hbar}{m}
            y\right)^2 + \omega^2 = 0.          
\label{mex4}
\end{equation}
So, solving the classical equation of motion (3) and equation (4) provides all
the information that determines the time-evolution of the quantum mechanical
WP (2).

The complex variable $\left( \frac{2\hbar}{m} y \right)$ of the Riccati equation will
play the key role in our construction of the CSs. Since, for the HO with TD frequency and the dissipative systems the Hamiltonian is no longer a constant of motion, a different quantity shall be used for the factorization method; namely, the Ermakov invariant. This
invariant can be obtained by eliminating the frequency $\omega$ from Eqs.(3)
and (4). Usually, for this purpose, Eq.(4) is written in the form of a (real,
NL) Ermakov equation that can be obtained by introducing a new variable
$\alpha (t)$ via 
\begin{equation}
\frac{2\hbar}{m}y_{I} = \frac{1}{\alpha^{2}(t)},	
\label{mex5}
\end{equation}
where $\alpha = \sqrt{\frac{2m}{\hbar}  \langle \widetilde{x}^2 \rangle }$ is directly proportional to the WP width. Inserting this definition (5) into the imaginary part of Eq.(4) yields
\begin{equation}
\frac{2\hbar}{m}y_{R} = \frac{\dot{\alpha}}{\alpha},
\label{mex6}
\end{equation}
and, finally, inserting (5) and (6) into the real part of (4) leads to the Ermakov equation 
\begin{equation}
\ddot\alpha + \omega^2(t) \alpha   = \frac{1}{\alpha^3}.
\label{mex7}
\end{equation}
Via the afore-mentioned elimination process, the invariant can be obtained as
\begin{equation}
\begin{array}{rl}
I_L& = \displaystyle\frac{1}{2} \left[ \left( \dot{\eta} \alpha - \dot{\alpha} \eta 
  \right)^2 + \left( \frac{\eta}{\alpha} \right)^2 \right] =  \frac{1}{2}
\alpha^{2}  \left[ \left( \dot{\eta}  - \frac{\dot{\alpha}}{\alpha} \eta 
  \right)^2 + \left( \frac{\eta}{\alpha^{2}} \right)^2 \right] \\[3ex]
       &= \displaystyle\frac{1}{2}
\alpha^{2}  \left[ \left( \dot{\eta}  - \left(\frac{2 \hbar}{m}y_{R}\right)\eta 
  \right)^2 + \left( \left( \frac{2 \hbar}{m}y_{I} \right) \eta \right)^2 \right] =  \textrm{const}.
\end{array}
\label{mex8}
\end{equation}
The corresponding quantum mechanical operator can be obtained easily by
replacing $\dot{\eta}$ by the momentum operator  $p_{op} = \frac{·\hbar}{i} \frac{\partial}{\partial x}$ divided by $m$ and $\eta$ by $x$, taking into account that
$p_{op}$ and $x$ do not commute.
 
\subsection{Complex Riccati equation and supersymmetric quantum mechanics}

There are different ways of treating the Riccati equation (4); in the following, we
will use a method that best shows the similarity to supersymmetric quantum
mechanics and point out the properties that have qualitative consequences for the
corresponding WP solutions. For this purpose we take advantage of the fact
that the inhomogeneous Riccati equation can be transformed into a homogeneous
Bernoulli equation if a particular solution $\left( \frac{2\hbar}{m} \widetilde{y}
\right)$ of the Riccati equation is known. The general solution of Eq.(4) is
then given by $\frac{2\hbar}{m} y  =  \frac{2\hbar}{m} \widetilde{y} +
\frac{2\hbar}{m} v(t)$ where $\frac{2\hbar}{m} v(t)$ fulfils
\begin{equation}
\left( \frac{2\hbar}{m}\dot{v} \right) + \left( \frac{4\hbar}{m}\widetilde{y} \right)
\left(\frac{2\hbar}{m}v \right)+ \left(\frac{2\hbar}{m}v\right)^{2} = 0.
\label{mex9}
\end{equation}
The coefficient $A = 2 \left( \frac{2\hbar}{m}\widetilde{y} \right)$ of the linear
term depends on the particular solution; further details will be discussed
later. The NL Bernoulli equation (9) can be linearized via $\frac{2\hbar}{m}v = \frac{1}{w(t)}$ to yield
\begin{equation}
\dot{w} - Aw=1,
\label{mex10}
\end{equation}
which has the solution
\begin{equation}
w(t)=\left[ w_{0} +  \int^{t}_{ } dt' e^{- \int^{t'} dt" A}  \right] e^{ \int^{t} dt' A}.
\label{mex11}
\end{equation}
Particularly for constant $A$, this reduces to
\begin{equation}
w(t) = \frac{1}{A} \left(e^{A t} - 1 \right) + w_{0} e^{A t},  
\label{mex12}
\end{equation}
depending on the (complex) initial value $w_{0} = \left(\frac{2\hbar}{m}v_{0}
\right)^{-1}$. Because Eqs.(4) and (9) are nonlinear equations, this
dependence on the initial conditions can be very sensitive and lead to
qualitatively quite different behaviour of the WP width. If, e.g., for the HO
with constant frequency $\omega = \omega_0$ the particular solution is chosen
to be $\frac{2\hbar}{m} \widetilde{y} = i \frac{2\hbar}{m} \widetilde{y}_{I} = i
\omega_0$ (a solution with a minus sign is mathematically also possible but
would results in an unphysical WP solution), i.e., $A = i 2 \omega_0$, a WP
with constant width $\alpha_0 = \omega_0^{-1/2} = \left( \frac{2m}{\hbar}
  \langle \widetilde{x}^2 \rangle _0 \right)^{1/2}$ is obtained. Whereas, for any choice
$\alpha_0 \neq \omega_0^{-1/2}$ corresponding to $\frac{2\hbar}{m}v_{0} =
\frac{1}{w_{0}} =  \left( \frac{1}{\alpha_0^2} - \omega_0  \right) \neq 0$, an oscillating width is obtained (see also \cite{22}).

The similarity to supersymmetric quantum mechanics becomes obvious when we
introduce the abbreviated form ${\mathcal I}(t) =   \int^{t}_{ } dt' e^{- \int^{t'}_{ } dt" A(t")}$, allowing for the rewriting of the general solution of Eq.(4) as
\begin{equation}
 \frac{2\hbar}{m} y(t) = \frac{2\hbar}{m} \widetilde{y} + \frac{d}{dt}  \textrm{ln} 
 [ w_{0} + {\cal I} (t) ],
\label{mex13}
\end{equation}
defining a one-parameter family of solutions depending on the initial value of
$w_0 = \left(\frac{2\hbar}{m}v_{0}
\right)^{-1}$ as parameter. Comparison with supersymmetric quantum mechanics \cite{18,19,20,23,24} shows that solution (13) is formally identical to the most general superpotential
\begin{equation}
\widetilde{W}(x) = W(x) + \frac{d}{dx}  \textrm{ln} 
 [ \lambda_{1} + {\cal I}_{1}(x) ], 
\label{mex14}
\end{equation}
leading to a one-parameter family of complex isospectral potentials
\begin{equation}
\widetilde{V}_{1}(x) = \widetilde{W}^{2} - \frac{d}{dx} \widetilde{W} = V_{1} - 2
\frac{d^{2}}{dx^{2}}  \textrm{ln} 
 [ \lambda_{1} + {\cal I}_{1}(x) ],
\label{mex15}
\end{equation}
that have the same supersymmetric partner potential $V_2(x)$ (see, e.g., \cite{24,25,26,27,28,29}). In this case, the integral ${\cal I}_{1}(x)$ is defined as
\begin{equation}
{\cal I}_{1}(x) =  \int^{x}_{-\infty} dx' \,  \Psi_{1}^{2}(x)
\label{mex16}
\end{equation}
where $\Psi_{1}(x)$ is the normalized ground state wave function of the SE
with potential $V_{1}(x) = W^{2}(x) - \frac{d}{dx} W(x)$ and $\lambda_{1}$ is a (usually real) parameter.

A major difference between this supersymmetric situation and the one in our
case (apart from replacing the spatial variable by a temporal one) is the fact
that the variables of the nonlinear Eqs.(4) and (9) are {\it complex}, whereas
$\widetilde{W}(x), W(x)$ and ${\cal I}_{1}(x)$ are real quantities; also the
parameter $w_0 $ in our case is generally complex. This provides a larger variety but, nevertheless, certain methods and results can be transferred from one system to the other (which will be discussed elsewhere).

\subsection{Generalized creation and annihilation operators and corresponding coherent states}

In the following it will be shown how the usual creation/annihilation operators
for the HO with constant frequency $\omega_0$ are related to the complex Riccati equation (4), how they can be generalized to take into account the additional dynamical aspects concerning the WP width and what consequences this has for the corresponding generalized coherent states.

In the usual treatment of the HO in terms of creation and annihilation operators, the Hamiltonian
\begin{equation}
H = \frac{1}{2m}  p_{op}^{2} + \frac{m}{2}  \omega_{0}^{2}  x^{2} = \hbar
\omega_{0} \left( a^{+} a  + \frac{1}{2} \right),
\label{mex17}
\end{equation}
with $p_{op} = \frac{\hbar}{i} \frac{\partial}{\partial x}$ is divided by $\hbar
\omega_{0}$ to yield
\begin{equation}
\widetilde{H} = \frac{H}{\hbar
\omega_{0}} = \left( a^{+} a  + \frac{1}{2} \right),
\label{mex18}
\end{equation}
where $\widetilde{H}$ is dimensionless (but $\frac{H}{\omega_{0}} = \alpha_0^2 H$ would have the dimension of an action) and the
creation and annihilation operators $ a^{+}$ and $a$ are defined as 
\be
\begin{array}{rl}
a &= i \displaystyle\sqrt{\frac{m}{2 \hbar \omega_{0}}} \left(
    \frac{p_{op}}{m} - i \omega_{0} x  \right)\\[3ex]
a^{+}&=-i \displaystyle\sqrt{\frac{m}{2 \hbar \omega_{0}}} \left(
    \frac{p_{op}}{m} + i \omega_{0} x  \right).
\end{array}
\label{mex19}
\ee
The corresponding CS is a WP with constant width, corresponding to the
particular solution of the Riccati equation with $\frac{2\hbar}{m} \widetilde{y} =
i  \frac{2\hbar}{m} \widetilde{y}_{I} = i  \frac{2\hbar}{m} y_{I} = i
 \omega_0$. Replacing $\omega_0$ by $\frac{1}{\alpha_0^2}$ or
 $\frac{2\hbar}{m} y_{I}$, respectively, the operators $a$ and $a^{+}$ can be rewritten as
\be
\begin{array}{rl}
a &= i \displaystyle\sqrt{\frac{m}{2 \hbar}} \alpha_{0} \left(
    \frac{p_{op}}{m} - i \left(\frac{2 \hbar}{m} y_{I}\right) x \right)\\[3ex]
a^{+}&=-i \displaystyle\sqrt{\frac{m}{2 \hbar}} \alpha_{0} \left(
    \frac{p_{op}}{m} + i \left(\frac{2 \hbar}{m} y_{I}\right) x \right).
\end{array}
\label{mex20}
\ee
As we have shown, already for $\omega =  \omega_0$ = const.,  solutions with TD
WP width exist, i.e., $\alpha_0$ turns into $\alpha (t)$ and $\dot{\alpha}
\neq 0$, hence $\frac{2\hbar}{m}y_{R} = \frac{\dot{\alpha}}{\alpha}$, must be
taken into account. Obviously, the same also applies for the HO with TD
frequency $\omega = \omega (t)$. Therefore, in Eqs.(20 a,b) $\alpha_{0}$ must be
replaced by $\alpha (t)$ and $i \left(\frac{2 \hbar}{m} y_{I}\right)$ by $\left(\frac{2 \hbar}{m} y \right)$, thus leading to 
\be
\begin{array}{rl}
\widetilde{a}(t) &= i \displaystyle\sqrt{\frac{m}{2 \hbar}} \alpha \left(
    \frac{p_{op}}{m} - \left(\frac{2 \hbar}{m} y \right) x \right)\\[3ex]
\widetilde{a}^{+}(t)&=-i \displaystyle\sqrt{\frac{m}{2 \hbar}} \alpha \left(
    \frac{p_{op}}{m} - \left(\frac{2 \hbar}{m} y^{\ast}\right) x \right).
\end{array}
\label{mex21}
\ee
It is easy to check that they satisfy the standard commutation relations if
Eq.(5) is fulfilled.

Since, at least for TD frequency $\omega$, the corresponding Hamiltonian is no
longer a constant of motion, one might ask if $\widetilde{a}(t)$ and $\widetilde{a}^{+}(t)$, as defined above, are constants of motion, i.e., if they fulfill
\begin{equation}
\frac{\partial}{\partial t} \widetilde{a} + \frac{1}{i \hbar} [ \widetilde{a}, H
]_{-} = 0
\label{mex22}
\end{equation}
(with $[~,~]_- =$ commutator). To answer this question, $x$, $p$ and hence $H$ shall be expressed in terms of
$\widetilde{a}(t)$ and $\widetilde{a}^{+}(t)$, leading to
\be
\begin{array}{rl}
x &= \displaystyle\frac{\alpha}{2} \sqrt{\frac{2 \hbar}{m}} ( \widetilde{a} + \widetilde{a}^{+} ),\\[3ex]
p &= m \displaystyle\frac{\alpha}{2} \sqrt{\frac{2 \hbar}{m}} \left\{ \left(\frac{2
      \hbar}{m} y^{\ast}\right) \widetilde{a} + \left(\frac{2 \hbar}{m} y
  \right) \widetilde{a}^{+} \right\},
\end{array}
\label{mex23}
\ee
with
\begin{equation}
H =  \frac{\hbar}{4}  \alpha^{2} \left[ \left(\frac{2 \hbar}{m} y_{R}
  \right) ( \widetilde{a} + \widetilde{a}^{+} )    + i \left(\frac{2 \hbar}{m} y_{I}
  \right) ( \widetilde{a}^{+} - \widetilde{a} )
\right]^{2} + \frac{\hbar}{4} \omega^2  \alpha^{2} \left( \widetilde{a}^{+} + \widetilde{a}
\right)^{2}.
\label{mex24}
\end{equation}
This finally leads to
\begin{equation}
\frac{\partial}{\partial t} \widetilde{a} + \frac{1}{i \hbar} [ \widetilde{a}, H
]_{-} = - i \frac{1}{\alpha^{2}} \widetilde{a} \neq 0.
\label{mex25}
\end{equation}
So, $\widetilde{a}(t)$ and $\widetilde{a}^{+}(t)$ are no constants of motion but can be turned into such by simply introducing a phase factor according to 
\be
\begin{array}{rl}
a(t) &= \displaystyle\widetilde{a}(t)  e^{i \int^{t}_{ } dt'\frac{1}{\alpha^{2} }}\\[2ex]
a^{+}(t) &= \displaystyle\widetilde{a}^{+}(t)  e^{-i  \int^{t}_{ } dt'\frac{1}{\alpha^{2} }}.
\end{array}
\label{mex26}
\ee
In order to elucidate the meaning of this phase factor we return to the
Riccati equation (4) and apply another treatment, namely, the linearization using the ansatz
\begin{equation}
 \left(\frac{2\hbar}{m}y\right) =  \frac{\dot{\lambda}}{\lambda},
\label{mex27}
\end{equation} 
with the complex quantity $\lambda = u + i v = \alpha e^{i \varphi}$, yielding the linear complex Newtonian equation
\begin{equation}
\ddot{\lambda} + \omega^2(t){\lambda}   = 0 
\label{mex28}
\end{equation}
for $\lambda (t)$, if (27) is inserted into Eq.(4).

It should be mentioned that generalized creation and annihilation operators
for the HO with TD frequency using a complex variable corresponding to
$\lambda$ had been used by Malkin et al. \cite{30}. Real and imaginary parts
of $\lambda$ and their time-derivatives are also the elements of the  $2
\times 2$ matrices of the real symplectic group $Sp(2,R)$ that provides the
representation of canonical transformations in TD quantum mechanics \cite{31}.

Using the polar form of $\lambda$ and inserting (27) into the Riccati equation (4), one obtains from the imaginary part a relation between the phase and amplitude of $\lambda$ as 
\begin{equation}
\dot\varphi = \frac{1}{\alpha^2}.
\label{mex29}
\end{equation}
So, the phase factor $\int^{t}_{ } dt'\frac{1}{\alpha^{2} }$ is (up to a
constant) just the angle $\varphi (t)$ in the complex plane of the quantity
$\lambda (t)$ that allows for the linearization of the Riccati equation (4). (For further details, see \cite{22,32}.)

Next, it shall be shown that one can use the definition of the generalized
annihilation operator $a (t)$ (Eq. 21a or 26a; the phase factor shall be
omitted in the following since it can be absorbed in the purely TD function
$K(t)$ in the exponent of the WP/CS) or into $N(t)$) to create a CS $ \vert z \rangle $ that is an eigenstate of $a(t)$ with complex eigenvalue $z$, i.e.,
\begin{equation}
a(t)  \vert z \rangle  = z  \vert z \rangle .
\label{mex30}
\end{equation}
First, the complex eigenvalue $z$ shall be determined in terms of $\langle x \rangle _{z} =
\eta$ and $\langle p \rangle _{z} = m \dot{\eta}$, where Eq.(30) is assumed to be valid. From
\be
\begin{array}{rl}
\langle x \rangle _{z} &= \displaystyle\sqrt{\frac{\hbar}{2 m}} \alpha (z^{\ast} + z ) = \sqrt{\frac{2 
    \hbar}{m}} \alpha z_{R} = \eta, \\[3ex]
\langle p \rangle _{z} &= \displaystyle\sqrt{\frac{\hbar}{2 m}} \alpha m \left[ \left(\frac{2 \hbar}{m} y
  \right) z^{\ast} +  \left(\frac{2
      \hbar}{m} y^{\ast}\right) z \right] = m \dot{\eta},
\end{array}
\label{mex31}
\ee
follows that
\be
\begin{array}{rl}
z_{R} &= \displaystyle\sqrt{\frac{m}{2
    \hbar}} \frac{\eta}{\alpha} = \frac{1}{\sqrt{2}}\sqrt{\frac{m}{\hbar}} \alpha \left(\frac{2 \hbar}{m} y_{I}\right) \eta,\\[3ex]
z_{I} &=\displaystyle \sqrt{\frac{m}{2
    \hbar}}  \left( \dot{\eta} \alpha - \eta \dot{\alpha}
\right) =  \frac{1}{\sqrt{2}}\sqrt{\frac{m}{\hbar}} \alpha \left[ \dot{\eta} -
  \left(\frac{2 \hbar}{m} y_{R}\right) \eta \right],
\end{array}
\label{mex32}
\ee
or
\begin{equation}
z = \sqrt{\frac{m}{2
    \hbar}} \left[ \left( \frac{\eta}{\alpha} \right)  + i \left( \dot{\eta} \alpha - \eta \dot{\alpha}
\right) \right] = \frac{1}{\sqrt{2}}\sqrt{\frac{m}{\hbar}} \alpha \left[
\left(\frac{2 \hbar}{m} y_{I}\right) \eta + i \left( \dot{\eta} - \left(\frac{2 \hbar}{m}
  y_{R}\right) \eta \right) \right].   
\label{mex33}
\end{equation}
This shows the connection between the eigenvalues $z$ (or $z^{\ast}$) and the Ermakov invariant as
\begin{equation}
I_{L} = \frac{\hbar}{m} \left( z_{I}^{2} + z_{R}^{2} \right) = \frac{\hbar}{m}
z z^{\ast} = \frac{\hbar}{m}  \vert z \vert^{2}.
\label{mex34}
\end{equation}
Immediately from this it follows that the operator corresponding to the
Ermakov invariant, when $p = m \dot{x}$ is replaced by the operator $p_{op} =
\frac{\hbar}{i} \frac{\partial}{\partial x}$ (and taking into account that
$[p,x]_- = \frac{\hbar}{i}$), can be written in terms of our generalized
creation and annihilation operators as $I_{L,op} = \frac{\hbar}{m} \left[
  a^{+}(t) a(t) + \frac{1}{2} \right]$, which is in agreement with the approach of Hartley and Ray \cite{11} to construct
these operators for the HO with TD frequency from $I_{L,op}$ via factorization.

In the position-space representation, the CS that is eigenstate of $a(t)$, 
\begin{equation}
\langle x \vert a(t) \vert z \rangle  = z \langle x \vert z \rangle  \quad \textrm{or } \quad  z  \Psi_{z}(x) = i
\sqrt{\frac{m}{2 \hbar}} \alpha  \left\{ \frac{\hbar}{mi} \frac{\partial}{\partial x} - \left(\frac{2 \hbar}{m} y
\right)x \right\} \Psi_{z}(x),
\label{mex35}
\end{equation}
can be given as
\begin{equation}
\Psi_{z}(x,t) = M(t) \exp \left\{ \frac{im}{2 \hbar} \left( \frac{2 \hbar}{m}y
    \right) (x - \langle x \rangle  )^{2}  + \frac{i}{\hbar} \langle p \rangle  x + \frac{i}{2 \hbar} \langle p \rangle 
    \langle x \rangle  + \frac{1}{2} \left( z^{2} +  \vert z \vert^{2} \right) \right\},
\label{mex36}
\end{equation}  
which is, for $M(t) = N(t)  e^{- \frac{1}{2} \left( z^{2} +  \vert z  \vert^{2} \right)}$
and $N(t) = \left( \frac{m}{\pi \hbar} \right)^{1/4} \left( \frac{1}{\lambda} \right)^{1/2}$, identical to the normalized Gaussian WP (2),
\begin{equation}
\Psi_{z}(x,t) = \left( \frac{m}{\pi \hbar} \right)^{1/4} \left(
  \frac{1}{\lambda} \right)^{1/2} \exp \left\{ i \left[ y \, \widetilde{x}^{2}+ \frac{1}{\hbar}
    \langle p \rangle  \widetilde{x} + \frac{1}{2 \hbar} \langle p \rangle 
    \langle x \rangle  \right] \right\} = \Psi_{WP}(x,t),
\label{mex37}
\end{equation}  
where, for the choice of a complex normalization factor $N(t)= \left( \frac{m}{\pi \hbar} \right)^{1/4} \left( \frac{1}{\lambda} \right)^{1/2}$, the TD function in the exponent has the form $K(t) = \frac{1}{2 \hbar} \langle p \rangle  \langle x \rangle $. Using the polar form of $\lambda (t)$, i.e., $\lambda = \alpha e^{i \varphi}$ with $\varphi =  \int^{t}_{ } dt'\frac{1}{\alpha^{2} }$, $N(t)$ can be written as $N(t)= \left( \frac{m}{\pi \hbar \alpha^2} \right)^{1/4} e^{- i \varphi /2} = \left( \frac{1}{2  \pi \langle \widetilde{x}^2 \rangle (t)} \right)^{1/4} e^{- i \frac12 \int^{t} dt' \frac{1}{\alpha^2}}$. For $\alpha^2 = \alpha_0^2= \omega_0^{-1}$, $N(t)$ turns into $N(t)= \left( \frac{m \omega_0}{\pi \hbar} \right)^{1/4}  e^{- i \omega_0 t/2}$ and contributes the ground state energy of the HO that corresponds to $\widetilde E= \frac{1}{2m} \langle \widetilde p^2 \rangle  + \frac{m}{2} \omega_0^2  \langle \widetilde x^2 \rangle  = \frac{\hbar}{2} \omega_0$ in the case of the WP with constant width. Here we obtain a generalization for $\alpha = \alpha (t)$. The phase factors $e^{\pm i  \int^{t} dt' \frac{1}{\alpha^2}}$ occurring in the creation/annihilation operators as defined in Eq. (26) can be absorbed into $N(t)$ in a similar way.

As mentioned in the introduction, the CS can also be defined as displaced vacuum state, i.e., 
\begin{equation}
 \vert z \rangle  = \exp \left\{ z a^{+}(t) - z^{\ast} a(t) \right\}   \vert 0 \rangle  = D(z)    \vert 0 \rangle .
\label{mex38}
\end{equation}
In the position-space representation, the vacuum state $\phi_{0}$ can be obtained via
\begin{equation}
\langle x \vert a(t) \vert 0 \rangle  = i
\sqrt{\frac{m}{2 \hbar}} \alpha  \left\{ \frac{\hbar}{mi} \frac{\partial}{\partial x} - \left(\frac{2 \hbar}{m} y
\right)x \right\} \phi_{0}(x) = 0
\label{mex39}
\end{equation}
as
\begin{equation}
\langle x \vert0 \rangle  = \phi_{0}(x,t) = N(t)  e^{i y(t) x^{2}}.
\label{mex40}
\end{equation}
Note that due to $y(t)$ the exponent is now complex, in particular via $i y_R
x^2$, the term $\frac{im}{2 \hbar}  \frac{\dot{\alpha}}{\alpha} x^2$ already occurs here naturally, whereas it must be introduced in the approach of Hartley and Ray \cite{11} via a unitary transformation.

Using the Baker--Campbell--Hausdorff formula, the CS can be written in the usual form
\begin{equation}
 \vert z \rangle  = D(z)  \vert 0 \rangle  = e^{- \frac{1}{2}  \vert z \vert} \exp(z a^{+}(t))  \vert 0 \rangle  = e^{- \frac{1}{2}  \vert z  \vert^{2}} \sum_{n=0}^{\infty} \frac{z^{n}
  (a^{+})^{n}}{n !}   \vert 0 \rangle ,
\label{mex41}
\end{equation}
now with $a^{+}(t)$ as defined in (21b) (or (26b)). Evaluating $e^{z a^{+}}$
with the help of $z$ and $ \vert z \vert^2$, as given in (33) and (34), leads again in
the position-space representation to $\Psi_{z}(x,t)$ as given in (37).

Finally, it will be shown that our CS also fulfills the Schr\"odinger--Robertson
uncertainty relation and how the uncertainties can be expressed in terms of $y_R$ and $y_I$ (since this form will also remain valid when dissipative effects are included, as will be explained in the next section).

In terms of $z$, $z^{\ast}$ and $\left(\frac{2 \hbar}{m} y
  \right)$ the mean value of $\langle x^2 \rangle $ can be written as
\begin{equation}
\langle x^2 \rangle _{z} = \frac{\hbar}{2 m} \alpha^{2} ( z^{\ast 2} + z^2 + 2  \vert z \vert^2 +
1 )
\label{mex42}
\end{equation}
leading (together with Eq.(31a)) to the mean square deviation of position
\begin{equation}
\langle \widetilde{x}^2 \rangle _{z} = \langle x^2 \rangle _{z} - \langle x \rangle _{z}^{2} = \frac{\hbar}{2 m} \alpha^{2}.
\label{mex43}
\end{equation}
In the same way, from 
\begin{equation}
\langle p^2 \rangle_{z} = \frac{\hbar m}{2} \alpha^{2} \left[ \left(\frac{2 \hbar}{m} y
  \right)^2  z^{\ast 2} +  \left(\frac{2
      \hbar}{m} y^{\ast}\right)^2 z^2 + \left \vert\frac{2 \hbar}{m} y\right \vert^2 \left( 2  \vert z \vert^2 +
1 \right) \right]
\label{mex44}
\end{equation}
and using Eq.(31b), one obtains
\begin{equation}
\langle \widetilde{p}^2 \rangle_{z} = \langle p^2\rangle_{z} - \langle p\rangle_{z}^{2} = \frac{\hbar m}{2} \alpha^{2} 
\left \vert\frac{2 \hbar}{m} y\right \vert^2 = \frac{\hbar m}{2} \left[ \left(\frac{2
      \hbar}{m} y_{R}
  \right)^2 +  \left(\frac{2
      \hbar}{m} y_{I} \right)^2  \right]
\label{mex45}
\end{equation}
and from
\begin{equation}
\langle [x,p]_{+}\rangle_z = \langle xp + px\rangle_z = \hbar \alpha^{2} \left[ \left(\frac{2 \hbar}{m} y
  \right)  z^{\ast 2} +  \left(\frac{2
      \hbar}{m} y^{\ast}\right) z^2 + \left(\frac{2 \hbar}{m} y_{R} \right)  \left( 2  \vert z \vert^2 +
1 \right) \right]
\label{mex46}
\end{equation}
with (31a) and (31b) the correlation uncertainty
\begin{equation}
\left\langle \frac{1}{2}[\widetilde{x},\widetilde{p}]_{+} \right\rangle_z = \left\langle \frac{1}{2}[x,p]_{+} \right\rangle_z - \langle x\rangle_{z}\langle p\rangle_{z} = \frac{\hbar}{2} \alpha^{2} \left(\frac{2 \hbar}{m} y_{R} \right).
\label{mex47}
\end{equation}
From
\begin{equation}
\langle \widetilde{x}^2\rangle_{z}
\langle \widetilde{p}^2\rangle_{z} - \left\langle \frac{1}{2}[\widetilde{x},\widetilde{p}]_{+}\right\rangle_z^{2} = \frac{\hbar^2}{4} \alpha^{4} \left[ \left(\frac{2
      \hbar}{m} y_{R}
  \right)^2 +  \left(\frac{2
      \hbar}{m} y_{I} \right)^2  \right] - \frac{\hbar^2}{4} \alpha^{4} \left(\frac{2
      \hbar}{m} y_{R}
  \right)^2 = \frac{\hbar^2}{4}
\label{mex48}
\end{equation}
follows that our CS also fulfills the Schr\"odinger--Robertson minimum uncertainty condition.

\section{Complex Riccati equations, Ermakov invariants and coherent states for dissipative systems}

There exist several approaches for describing dissipative quantum systems using
modified effective one-body SEs where the effect of the environment on the
observable system is taken into account by dissipative friction terms without
considering the individual degrees of freedom of the environment. This usually
leads to NLSEs, SEs with explicitly TD Hamiltonians or non-Hermitian
Hamiltonians. Several of these approaches are discussed in the literature
\cite{12,13,14,15,16}. In this context it was also
investigated \cite{17} to ascertain if Ermakov invariants also exist for these
approaches. There are two NLSEs \cite{15,16} and two explicitly TD
Hamiltonians \cite{12,33} for which this is the case and that will be
discussed subsequently. It will also be shown that they are not independent of
each other. The NLSEs are related in a way that is connected with a (unitary)
phase transformation (for further details, see \cite{34}; the NLSEs and the TD
Hamiltonians, however, are connected via a non-unitary transformation \cite{35}. Taking these facts into account, the Ermakov invariants of all four approaches are equivalent and can be transformed one into the other.

In the following, these approaches will be presented whilst giving the corresponding WP solutions, equation of motion for the classical trajectory and the complex Riccati equation describing the respective time-evolution of the WP maximum and width, as well as the resulting Ermakov invariants. The corresponding creation/annihilation operators for the (physically relevant) NLSE will be defined and the CSs for the dissipative systems obtained on this basis. Finally comparisons will be made with the case discussed previously.

\subsection{Explicitly time-dependent Hamiltonian of Caldirola and Kanai}

This approach \cite{12} starts on the classical level with an explicitly TD Lagrangian, 
\begin{equation}
 \hat{L}_{CK} = \left[ \frac{m}{2} \dot{x}^{2} - V \left( x \right) \right] e^{\gamma t}.
\label{mex49}
\end{equation}
With the corresponding canonical momentum
\begin{equation}
\hat{p} =  \frac{\partial}{\partial
  \dot{x}} \hat{L}_{CK} =
m \dot{x}   e^{\gamma t} =  p  e^{\gamma t},
\label{mex50}
\end{equation}
the Hamiltonian can be formulated as
\begin{equation}
\hat{H}_{CK} = \frac{1}{2m} e^{-\gamma t}  \hat{p}^{2} + e^{\gamma t} V \left(
 x \right),
\label{mex51}
\end{equation}
which yields the proper equation of motion including a linear momentum (or velocity) dependent friction force with friction coefficient $\gamma$,
\begin{equation} 
\dot{p} + \gamma p + \frac{\partial}{\partial x} V = m \ddot{x} + m \gamma 
\dot{x} + \frac{\partial}{\partial x} V = 0.
\label{mex52}
\end{equation}
But, $\hat{H}_{CK}$ neither represents the energy of the system, nor is it a constant of motion.

It is important to realize that not only the canonical momentum $\hat{p}$ is
different from the physical (kinetic) momentum $p = m \dot{x}$ but, in
particular, the transition from the {\it physical} variables $x$ and $p = m
\dot{x}$ to the {\it canonical} variables $\hat{x} = x$, $\hat{p} =
p e^{\gamma t}$ represents a {\it non-canonical} transformation.

The transition to quantum mechanics is achieved by replacing, as usual, the
{\it canonical} momentum with a differential operator according to $\hat{p}
\rightarrow \hat{p}_ {op} = \frac{\hbar}{i} \frac{\partial}{\partial
  x}$, leading to the corresponding Hamiltonian operator and, thus, to the modified SE
\be
\begin{array}{rl}
\displaystyle i \hbar\frac{\partial}{\partial t} \hat{\Psi}_{CK}
 \left(x,t\right)&=\hat{H}_{CK, op} \hat{\Psi}_{CK}\left( x,t \right)\\[3ex]
 &= \displaystyle\left\{e^{-\gamma t}\left(-\frac{\hbar^{2}}{2m}\frac{\partial^{2}}{\partial
 x^{2} } \right) + e^{\gamma t} V \right\}  \hat{\Psi}_{CK} \left( x, t \right).
\end{array}
\label{mex53}
\ee
For the systems considered in this work, this equation also possesses exact Gaussian WP solutions like (2). The equation of motion for the WP maximum is just one for the classical trajectory (including the friction force), in our notation
\begin{equation}
 \ddot \eta + \gamma \dot{\eta} + \omega^2 \eta = 0.
\label{mex54}
\end{equation}
The modified Riccati equation for the complex variable $\left( \frac{2\hbar}{m}
            \hat{y} \right)_{CK}$ reads
\begin{equation}  
\left( \frac{2\hbar}{m} \dot{\hat{y}}\right)_{CK} +  e^{- \gamma t}\left( \frac{2\hbar}{m}
            \hat{y} \right)^2_{CK}+  \omega^2   e^{\gamma t} =  0.  
\label{mex55}
\end{equation}
The imaginary part of this variable is again connected with the WP width via
$\left( \frac{2\hbar}{m}  \hat{y}_I \right)_{CK} =  \frac{\hbar}{2m
  \langle \widetilde{x}^2\rangle_{CK}}$, where the subscript CK denotes the mean value being
calculated with $\hat{\Psi}_{CK}$, the solution of Eq.(53). Introducing, like
in the conservative case, a new (real) variable $\alpha_{CK}$ via $\left(
  \frac{2\hbar}{m}  \hat{y}_I \right)_{CK} = \frac{1}{\alpha_{CK}^2}$, again allows for the transformation of this Riccati equation into a (real) Ermakov-type equation,
\begin{equation}
 \ddot \alpha_{CK} + \gamma \dot \alpha_{CK} + \omega^2 \alpha_{CK} =
 \frac{e^{- 2 \gamma t}}{\alpha_{CK}^3}.
\label{mex56}
\end{equation}
This equation, together with Eq.(56) for $\eta$, forms the required system of equations that possesses an exact Ermakov-type invariant, here given in the form
\begin{equation}
 \hat{I}_{CK} = \frac{1}{2} \bigg[  e^{2 \gamma t} \bigg( \dot \eta  \alpha_{CK} - \eta  \dot
            \alpha_{CK} \bigg)^2 +
            \bigg(\frac{\eta}{\alpha_{CK} }\bigg)^2\bigg] = \textrm{const}.
\label{mex57}
\end{equation}
A major point of criticism concerning this approach is an apparent violation of the uncertainty principle. Defining the uncertainty product via
\begin{equation}
U_{CK} = \langle \widetilde{x}^2\rangle_{CK}
\langle \widetilde{p}^2\rangle_{CK} = \langle \widetilde{x}^2\rangle_{CK}
\langle \widetilde{\hat{p}}^2\rangle_{CK} e^{-2 \gamma t}
\label{mex58}
\end{equation}
by expressing the physical momentum $p$ in terms of the canonical momentum
$\hat{p}$ as $p = e^{- \gamma t} \hat{p}$, yields an exponential decay of
this product. So, $U_{CK} $ can become smaller than $\hbar^2/4 $ and even
vanish for $t \rightarrow \infty$. (A solution of this problem will be given later.)

\subsection{Description in exponentially expanding coordinates}

This problem does not occur when applying a different {\it non-canonical}
transformation \cite{27} using an exponentially-expanding coordinate $Q =  e^{\gamma t/2}  x$ and
the corresponding canonical momentum $P = m \dot{Q} = m  e^{\gamma
  t/2} ( \dot{x}  +  \frac{\gamma}{2} x )$.

The corresponding Hamiltonian function
\begin{equation}  
\hat{H}_{exp} =   \frac{1}{2m} P^2  + \frac{m}{2}  \Omega^2  Q^2  =  \textrm{const}.
\label{mex59}
\end{equation}
with $\Omega^{2} =  \left( \omega^2 - \frac{\gamma^2}{4} \right)$ is not only a constant of motion but, expressed in terms of the physical variables $x$ and $p$ as 
\begin{equation}  
\hat{H}_{exp} =  \left[ \frac{1}{2m} p^2  + \frac{\gamma}{2} p x + \frac{m}{2}
  \omega^2  x^2 \right] e^{\gamma
  t} = \frac{1}{2m} p^2_{0} + \frac{m}{2}
  \omega^2  x^2_{0} = E_{0}
\label{mex60}
\end{equation}
shows that it also represents the initial energy $E_{0}$ of the system (with
$p_{0}$ and $x_{0}$ being the initial momentum and position).

From the Hamiltonian equations of motion one obtains the equivalent Newtonian equation
\begin{equation}    
 \ddot Q + \left( \omega^2 - \frac{\gamma^2}{4} \right) Q = 0  
\label{mex61}
\end{equation}
which, expressed in terms of the physical variable $x$, again provides the desired equation of motion including the friction force.

Canonical quantization according to $Q \rightarrow Q_{op} = Q,  P \rightarrow
P_{op} = \frac{\hbar}{i} \frac{\partial}{\partial Q}$ provides the corresponding SE with
Gaussian WP solutions, only now the physical variable $x$ is replaced by the
canonical variable $Q$ (and $\omega^2$  by $\Omega^2$). Apart from this, it looks exactly like the SE in the conservative case.

Obviously, the corresponding Riccati and Ermakov equations can immediately be written as
\begin{equation}
\left( \frac{2\hbar}{m} \dot{\hat{y}} \right)_{exp} + \left(\frac{2\hbar}{m}
            \hat{y} \right)_{exp}^2 + \Omega^2 = 0       
\label{mex62}
\end{equation}
and
\begin{equation}
\ddot\alpha_{exp}+ \Omega^2(t) \alpha_{exp}  = \frac{1}{\alpha_{exp}^3}.
\label{mex63}
\end{equation}
Expressed in terms of $\alpha_{exp}$ and $Q$, the corresponding Ermakov invariant can be written in the form
\begin{equation}
\hat{I}_{exp} = \frac{1}{2}  \alpha_{exp}^2 \left[ \left( \dot{Q} -  \frac{\dot{\alpha}_{exp}}{\alpha_{exp}} Q 
  \right)^2 + \left( \frac{1}{\alpha_{exp}^2} Q  \right)^2 \right] =  \textrm{const}.  
\label{mex64}
\end{equation}
which is not only independent of $\omega$, but also does not explicitly
contain the friction coefficient $\gamma$ (and therefore also exists for
$\gamma = \gamma (t)$).

This approach is related to the afore-mentioned one by Caldirola and Kanai via
a {\it unitary} transformation (on the classical level), $Q = e^{\gamma
  t/2}  \hat{x},  P = \hat{p}   e^{- \gamma t/2}    + 
m \frac{\gamma}{2} \hat{x}  e^{\gamma t/2}$, and the Hamiltonians and action functions are connected via 
\begin{equation}
\hat{H}_{exp} = \hat{H}_{CK} + \frac{\partial}{\partial t} \hat{F}_2            
\label{mex65}
\end{equation}
with
\begin{equation}
\hat{F}_2(\hat{x},P,t) = \hat{x} P e^{\gamma t/2} - m \frac{\gamma}{4}
\hat{x}^2   e^{\gamma t}
\label{mex66}
\end{equation}
and
\begin{equation}
\hat{S}_{exp} = \hat{S}_{CK} + m \frac{\gamma}{4}
\hat{x}^2   e^{\gamma t} = \left( S +  m \frac{\gamma}{4}
x^2 \right) e^{\gamma t}. 
\label{mex67}
\end{equation}
Note that with Schr\"odinger's definition of the wave function $\Psi$ via the
action function, $S_c = \frac{\hbar}{i}  \textrm{ln} \Psi$, this canonical
transformation corresponds to a (unitary) phase factor $\frac{im}{\hbar}  \frac{\gamma}{4} x^2$ in the wave function.

The quantized version of this approach causes no problem with the uncertainty
principle \cite{33} providing one consistently translates operators {\it and} wave
function from the canonical level to the physical one. How this can be
achieved becomes obvious after we present the approaches that already start on
the quantum mechanical level and modify the Hamiltonian {\it operator} to include the effect of the dissipative environment, leading to NLSEs.

\subsection{NLSE with complex logarithmic nonlinearity}

The first attempt to introduce dissipation into the SE via a (real)
logarithmic nonlinearity by Kostin \cite{13} was based on obtaining the correct Newtonian
equation including the friction force for the mean values. It however suffered
several shortcomings \cite{15,16}, particularly, the reversible
continuity equation for the probability density
$\varrho\left(x,t\right)=\Psi^{\ast}\left(x,t\right)\Psi\left(x,t\right)$ in
the case of an irreversible dissipative system lacked reasonable explanation. Our approach \cite{16} was therefore motivated by the attempt to break the time-reversal symmetry on all levels to include irreversibility, a phenomenon usually associated also with dissipative systems (but not necessarily occurring only in those systems). This can be achieved by introducing a diffusion term into the continuity
equation thus turning it into a Fokker--Planck-type equation (in particular a
Smoluchowski equation). Following a method by Madelung and Mrowka{36},
the (real) Smoluchowski equation can be separated into two complex equations,
namely a (modified) SE for the wave function $\Psi\left(x,t\right)$ and its complex conjugate,
$\Psi^{\ast}\left(x,t\right)$, if the separation condition
\begin{equation}  
-D \frac{\frac{\partial^{2}}{\partial x^{2}}\varrho}{\varrho}
  = \gamma\left(\textrm{ln}  \varrho -\langle \textrm{ln}  \varrho\rangle\right)
\label{mex68}
\end{equation}
with diffusion coefficient $D$ is fulfilled. This leads to the NLSE
\begin{equation}
 i\hbar \frac{\partial}{\partial t} \Psi_{NL}(x,t) = \left\{ -\frac{\hbar^{2}}{2m} \frac{\partial^{2}}{\partial x^{2}} + V(x) +  \gamma \frac{\hbar}{i} \left(\textrm{ln}  \Psi_{NL} -\langle  \textrm{ln}  \Psi_{NL}\rangle
 \right) \right\} \Psi_{NL}(x,t)
\label{mex69}
\end{equation}
with a complex logarithmic nonlinearity (for details, see e.g.,
\cite{16}). The real part of this nonlinearity depends on the phase of the WP,
is identical to Kostin`s term and influences the motion of the WP maximum. The imaginary part that corresponds to the diffusion term in the Smoluchowski equation has no influence on the WP maximum, but a strong one on the dynamics of the WP width.

The equation of motion for the WP maximum is again identical to Eq.(54). The equation for the time-dependence of the width can now be obtained from the modified complex Riccati equation
\begin{equation}  
 \left(\frac{2\hbar}{m}\dot{y} \right)_{NL} + \gamma
 \left(\frac{2\hbar}{m}y\right)_{NL}+ \left(\frac{2\hbar}{m}y\right)^{2}_{NL}+\omega^{2}(t) = 0.
\label{mex70}
\end{equation}
Again, the imaginary part has the same relation to the WP width (or position
uncertainty) as in the conservative case, $\left( \frac{2\hbar}{m}  y_I \right)_{NL} =  \frac{\hbar}{2m
  \langle \widetilde{x}^2\rangle_{NL}}$.  Introducing the variable $\alpha_{NL}$ via $\left(
  \frac{2\hbar}{m}  y_I \right)_{NL} = \frac{1}{\alpha_{NL}^2}$, the real part of the complex Riccati-variable now takes the modified form
\begin{equation} 
   \frac{2\hbar}{m}  y_{R,NL} = \frac{\dot\alpha_{NL}}{\alpha_{NL}} - \frac{\gamma}{2}.
\label{mex71}
\end{equation}
With the help of these relations, the Riccati equation now turns into the Ermakov-type equation
\begin{equation} 
 \ddot \alpha_{NL} + \left( \omega^2 - \frac{\gamma^2}{4} \right) \alpha_{NL} = \frac{1}{\alpha^3_{NL}},
\label{mex72}
\end{equation}
i.e., exactly the same equation as for $\alpha_{exp}$ in the expanding system. Together with the dissipative equation for the WP maximum, Eq.(54), this leads to the exact Ermakov invariant
\begin{equation} 
 I_{NL} =\frac12  e^{\gamma t}   \alpha_{NL}^2 \left[ \bigg( \dot \eta -
 \left( \frac{\dot \alpha_{NL}} {\alpha_{NL}} - \frac{\gamma}{2} \right) \eta \bigg)^2 
+ \left( \frac{1}{\alpha_{NL}^2}  \eta \right)^2 \right] =  \mbox{const}.
\label{mex73}
\end{equation}

\subsection{General nonlinear  ``friction potential''}

A different attempt at finding a dissipative modification of the SE was used
by S\"ussmann, Albrecht{14} and Hasse{15} in trying to add possible
combinations of position and momentum operators and their mean values to the
linear SE in order to obtain a modification that provides the correct equation
of motion (54) for the mean value of position, including the linear friction
force. For this purpose they added some kind of ``friction potential'' $W(x,p;\Psi)$ to the linear SE, where $W$ introduced a nonlinearity due to the occurrence of the mean values.

The general form of this term is 
\begin{equation}
W_{Gen} = \gamma \langle p\rangle ( x - \langle x\rangle ) + C \frac{\gamma}{2} \left[ ( x - \langle x\rangle ), ( p -
\langle p\rangle )  \right]_+
\label{mex74}
\end{equation}
(with $[~,~]_+ =$ anticommutator) that, for any choice of $C$, leads to the desired equation of motion for $\langle x\rangle = \eta (t)$. Particular choices for $C$ lead to the following friction terms:
\begin{eqnarray}
\textrm{S\"ussmann (C =1):} & W_{\textit{S\"uss}} = \frac{\gamma}{2} \left[ ( x - \langle x\rangle ), p  \right]_+
\label{mex75}\\[2ex]
\textrm{Albrecht (C =0):} & W_{Alb} = \gamma \langle p\rangle ( x - \langle x\rangle )
\label{mex76}\\[2ex]
\textrm{Hasse (C = 1/2) :} & W_{Has} = \frac{\gamma}{4} \left[ ( x - \langle x\rangle ), (p  +  \langle p\rangle )  \right]_+
\label{mex77}
\end{eqnarray}
Whereas the choices for $C=1$ and $C=0$ lead to a wrong reduced frequency for
the damped HO, $C = \frac{1}{2}$ provides the correct one.

It should be mentioned that $W_{Has}$ can be written as a combination of the other two approaches as
\begin{equation}
W_{Has} = \frac{1}{2}\left( W_{\textit{S\"uss}} +  W_{Alb} \right).
\label{mex78}
\end{equation}
The NLSE of Hasse also has Gaussian WP solutions for the systems under
consideration and the equations of motion for its maximum and width (and
therefore also the corresponding Ermakov invariant) are identical to those of
our logarithmic NLSE (69). There is a small point that cannot be explained
sufficiently in Hasse's approach, namely, a contribution to the energy
originating from the non-vanishing mean value of $W_{Has}$, $\langle W_{Has}\rangle \neq 0$.

Comparison of $W_{Has} \Psi_{WP} (x,t)$ with our logarithmic term $W_{SCH} \Psi_{WP} (x,t)$ shows that
\begin{equation}
 W_{SCH} = W_{Has} - \langle W_{Has}\rangle,
\label{mex79}
\end{equation}
so, in our case, the non-vanishing mean value does not occur. The relationship
between this approach using the ``friction potential'' and our logarithmic
ansatz is connected with a unitary transformation and is discussed in more detail in \cite{34}.

In the following, the logarithmic approach will be considered explicitly since
the relevant part for the CS is essentially the complex Riccati equation (70)
that is identical to Hasse's approach. But it is more straightforward to
bridge the gap between the afore-mentioned explicitly TD Hamiltonians based on
non-canonical modifications of classical mechanics and the nonlinear modifications of the SE by using the complex logarithmic term.

\subsection{Connection between the canonical and nonlinear approaches}

To establish the connection between the explicitly TD Caldirola--Kanai approach and the logarithmic NLSE (69) we refer to Schr\"odinger's first communication on wave mechanics \cite{1} where he starts from the Hamilton--Jacobi equation
\begin{equation} 
\frac{\partial}{\partial t} S  +H \left( x, \frac{\partial}{\partial x}
S, t \right) = 0
\label{mex80}
\end{equation}
with the action function $S$ and the momentum $p = \frac{\partial}{\partial
  x} S$. He introduced the wave function $\Psi(x,t)$ via $S_{c} =  \frac{
    \hbar}{i}  \textrm{ln} \Psi$, where the subscript $c$ (added by us) indicates
that this action is a complex quantity, since $\Psi$ is, in general, a complex
function (a fact that Schr\"odinger did not like at all in the beginning \cite{37}). Via a variational ansatz, Schr\"odinger arrived at the Hamiltonian
operator $H_{L} = - \frac{\hbar^{2}}{2m}
\frac{\partial^{2}}{\partial x^{2}} + V \left( x \right)$.

We now reverse Schr\"odinger's procedure, starting with Eq.(69) (divided by $\Psi$, which causes no problems for Gaussian WPs) and, using the definition of $S_{c}$, arrive at 
\begin{equation} 
\left( \frac{\partial}{\partial t}+\gamma \right) S_{c} + H = -\gamma\langle S_{c}\rangle.
\label{mex81}
\end{equation}
This is, of course, as little rigorous as Schr\"odinger's first attempt
was. However, it follows his idea of connecting the classical Hamilton--Jacobi
theory with a wave (mechanical) equation.  The purely TD term $-\gamma\langle S_{c}\rangle$ is necessary mainly for normalization purposes (can therefore be absorbed by the normalization coefficient) and shall be neglected in the following.

Multiplying the remaining Eq.(81) by $e^{\gamma t}$ and using the definitions
\begin{equation} 
 \hat{S}_{c} = e^{\gamma t}S_{c}   \quad \mbox{and}  \quad \hat{H} = e^{\gamma t}H
\label{mex82}
\end{equation}
it can be rewritten as Hamilton--Jacobi equation
\begin{equation} 
\frac{\partial}{\partial t} \hat{S}_{c}+\hat{H} = 0.
\label{mex83}
\end{equation}
From the definition of the action function, it follows that the wave function
$\hat{\Psi} (x,t)$ in the transformed (canonical) system is connected with the wave function
$\Psi (x,t)$ in the physical system via the {\em non-unitary} relation 
\begin{equation} 
 \textrm{ln} \hat{\Psi} = e^{\gamma t}  \textrm{ln} \Psi.
\label{mex84}
\end{equation}
Consequently, the (complex) momenta in the two systems are connected via
\begin{equation} 
\hat{p}_{c} = \frac{\hbar}{i} \frac{\partial}{\partial x} \textrm{ln} \hat
 {\Psi} = e^{\gamma t} \frac{\hbar}{i} \frac{\partial}{\partial x} 
 \textrm{ln} \Psi_{NL} = e^{\gamma t}p_{c},
\label{mex85}
\end{equation}
which is equivalent to the connection between the canonical and the kinetic momentum in the Caldirola--Kanai approach. The {\em non-canonical} connection between the classical variables $(x,p)$ and $(\hat{x},\hat{p})$ corresponds to the {\em non-unitary} transformation between $\Psi$ and $\hat{\Psi}$. 

Note:  Although $\Psi$ and $\hat{\Psi}$ depend explicitly on the same variables, $x$ and $t$, the two wave functions are analytically different functions of $x$ and $t$ and have different physical meanings due to the non-unitary transformation (84). Ignoring this fact leads to the apparent unphysical results like violation of the uncertainty principle (for further details, see \cite{34}).

Expressing $\hat{H}$ in terms of the canonical momentum $\hat{p}_{c}$ and following Schr\"odinger's quantization procedure, finally yields the modified SE (53) of the Caldirola--Kanai approach.

Because the WP solution of Eq.(53), $\hat{\Psi}_{CK} (x,t)$, is related to the WP solution $\Psi_{NL} (x,t)$ of the NLSE (69) via (84), this means particularly for the corresponding Riccati and Ermakov equations that
\begin{equation}
\left( \frac{2\hbar}{m} \hat{y} \right)_{CK} = e^{\gamma t}\left( \frac{2\hbar}{m}
            y \right)_{NL}
\label{mex86}
\end{equation}
and
\begin{equation} 
\alpha_{CK}  =  e^{- \gamma t/2} \alpha_{NL}.
\label{mex87}
\end{equation}
Taking this into account, Eqs.(55) and (56) turn into Eqs.(70) and (72) and
the Ermakov invariant $\hat{I}_{CK}$ turns exactly into $I_{NL}$.

\subsection{Uncertainties, bifurcations, creation and annihilation operators and coherent states for dissipative systems}

With the help of the WP solution of Eq.(69) the uncertainties of position and
momentum, their correlation and product as well as their contribution to the
energy of the system can be expressed in terms of $\alpha_{NL}$ and
$\dot{\alpha}_{NL}$ or $\left( \frac{2\hbar}{m} y_{I} \right)_{NL}$ and
$\left( \frac{2\hbar}{m} y_{R} \right)_{NL}$, respectively, where the latter
ones are more general since this form can be used identically for the
corresponding quantities in the case without dissipation (therefore the
subscript $NL$ is omitted). One obtains
\begin{equation}  
\langle \widetilde x^2\rangle_{NL} = \frac{\hbar}{2m} \alpha^2_{NL} = \frac{\hbar}{2m} 
\left( \frac{2\hbar}{m} y_{I} \right)^{-1} 
\label{mex88}
\end{equation}

\begin{equation}
\begin{array}{c}
\langle \widetilde p^2\rangle_{NL} = \frac{\hbar m}{2} \bigg[\left( \dot \alpha_{NL} - \frac{\gamma}{2}
           \alpha_{NL} \right)^2 + \frac{1}{\alpha^2_{NL}}
         \bigg] = \frac{\hbar m}{2}   \left( \frac{2\hbar}{m} y_{I}
         \right)^{-1} \left[ \left( \frac{2\hbar}{m} y_{R}
           \right)^{2} + \left( \frac{2\hbar}{m} y_{I} \right)^{2} \right], 
\end{array}
\label{mex89}
\end{equation}

\begin{equation} 
\langle [\widetilde x, \widetilde p]_+\rangle_{NL} = \hbar \left( \dot \alpha_{NL}  
        \, \alpha_{NL} - \frac{\gamma}{2} \alpha^2_{NL} \right) = \hbar \left(
        \frac{2\hbar}{m} y_{I} \right)^{-1} \left( \frac{2\hbar}{m} y_{R}
           \right), 
\label{mex90}
\end{equation}

\be
\begin{array}{c}
U_{NL} = \langle \widetilde x^2\rangle_{NL} \, \langle \widetilde p^2\rangle_{NL} = \frac{\hbar}{4} \left[
  1 + \left(  \dot \alpha_{NL}  \, 
        \alpha_{NL} - \frac{\gamma}{2} \alpha^2_{NL} \right)^{2} \right]
        =\frac{\hbar}{4} \left[
  1 + \left( \frac{2\hbar}{m} y_{I} \right)^{-1} \left( \frac{2\hbar}{m} y_{R}
           \right)^{2} \right]
\end{array}
\label{mex91}
\ee

\be
\begin{array}{rl}
 \widetilde E_{NL}&= \frac{1}{2m} \displaystyle\langle \widetilde p^2\rangle_{NL} + \frac{m}{2} \omega^2
        \langle \widetilde x^2\rangle_{NL}  
= \frac{\hbar}{4} \bigg\{ \left( \dot\alpha_{NL} -
          \frac{\gamma}{2} \alpha_{NL} \right)^2 + \frac{1}{\alpha^2_{NL}}
        + \omega^2 \alpha^2_{NL} \bigg\}\\[3ex]
&= \displaystyle\frac{\hbar}{4} 
\label{mex92} \left( \frac{2\hbar}{m} y_{I}
         \right)^{-1} \left[ \left( \frac{2\hbar}{m} y_{R}
           \right)^{2} + \left( \frac{2\hbar}{m} y_{I}
           \right)^{2} + \omega^{2} \right]. 
\end{array}
\ee
At this point we wish to recall the solution of the Riccati equation (4) via
transformation into a Bernoulli equation (9) with the help of a particular
solution of the Riccati equation, where this solution $\left( \frac{2\hbar}{m}
  \widetilde{y} \right)$ occurred in the Bernoulli equation as parameter $A = 2 \left( \frac{2\hbar}{m}
  \widetilde{y} \right)$. Including the dissipative term in Riccati equation (70), this
  parameter (already in the simples cases for constant $\left( \frac{2\hbar}{m}
  \widetilde{y} \right)$) now takes the form
\begin{equation}
A = \pm 2 \left( \frac{\gamma^{2}}{4} - \omega^{2} \right)^{1/2} 
\label{mex93}
\end{equation}
So, already for the free motion $(\omega = 0)$, the two values $A_\pm = \pm
\gamma$ are possible, leading to two different solutions with different (TD)
uncertainties and contributions to the energy, $\widetilde E_{NL \pm}$, as can be
seen from Eqs.(88-92). This kind of Hopf bifurcation also occurs for $\omega
\neq 0$ and can be expected due to the quadratic nonlinearity of the Riccati
and Bernoulli equations; however, it is not immediately obvious when
considering the equivalent Ermakov equations. Bifurcations of this kind are
common in classical nonlinear dynamics \cite{38}, but usually in connection
with real evolution equations. In our quantum mechanical context, the
bifurcation does not show up in the conservative case because either the two
solutions are degenerate $(V =0: A = \pm 0 )$ or one of the two leads to
unphysical results $( HO: A = - i 2 \omega_0 )$. Taking into account the dissipative environment changes this situation qualitatively. Different solutions of the Riccati equation (70) also entail different creation/annihilation operators for the respective system.

In the conservative case we found that the creation and annihilation operators are related to the Ermakov invariant via 
\begin{equation}
I_{L,op} = \frac{\hbar}{m} \left[ a^{+}(t)  a(t) + \frac{1}{2} \right] \quad \textrm{or }    
\quad I_{L} = \frac{\hbar}{m} 
z z^{\ast},
\label{mex94}
\end{equation}
i.e., they could also be obtained by factorization of $I_{L,op}$. Trying to do the same with $I_{NL}$ leads to
\be
\begin{array}{rl}
z_{NL} &= \displaystyle\sqrt{\frac{m}{2
    \hbar}} \left[ \left( \frac{\eta}{\alpha_{NL}} \right)  + i \left(
    \dot{\eta}  \alpha_{NL} -  \left( \dot{\alpha}_{NL} - \frac{\gamma}{2} \alpha_{NL} \right) \eta \right) \right] e^{\gamma t/2},\\[3ex]
z_{NL}^{\ast} &= \displaystyle\sqrt{\frac{m}{2
    \hbar}} \left[ \left( \frac{\eta}{\alpha_{NL}} \right)  - i \left(
    \dot{\eta}  \alpha_{NL} -  \left( \dot{\alpha}_{NL} - \frac{\gamma}{2} \alpha_{NL} \right) \eta\right) \right] e^{\gamma t/2},
\end{array}
\label{mex95}
\ee
or
\be
\begin{array}{rl}
a_{NL}(t) &= i \displaystyle\sqrt{\frac{m}{2 \hbar}} \alpha_{NL} \left(
    \frac{p_{op}}{m} - \left(\frac{2 \hbar}{m} y \right)_{NL} x  \right) 
  e^{\gamma t/2},\\[4ex]
a^{+}_{NL}(t)&=-i \displaystyle\sqrt{\frac{m}{2 \hbar}} \alpha_{NL} \left(
    \frac{p_{op}}{m} - \left(\frac{2 \hbar}{m} y^{\ast}\right)_ {NL} x
     \right)  e^{\gamma t/2}.
\end{array}
\label{mex96}
\ee
From (96a,b) it becomes obvious that, apart from the exponential factor $e^{\gamma t/2}$, the form of the creation and annihilation operators is identical to the ones defined in (21a,b). Only the solutions of Riccati equation (4) must be replaced by the ones of Eq.(70) belonging to the logarithmic NLSE (69).

Still, the meaning of the exponential factor $e^{\gamma t/2}$ needs to be
explained. Since $I_{NL}$ is a constant of motion for the dissipative system,
like $\hat{H}_{exp}$ on the canonical level, but the nonlinear Hamiltonian
and its mean value are not, the invariant should correspond to the canonical level
that is related to the physical one via the non-unitary transformation
$\textrm{ln} \hat{\Psi} = e^{\gamma t}  \textrm{ln} \Psi$. Therefore, the
factor $e^{\gamma t}$ in $I_{NL}$ should be omitted if it is used for the definition of the creation and annihilation operators that supply the wave functions and CSs which are solutions of the Hamiltonian operator on the physical level.

So, finally the generalized creation and annihilation operators for the dissipative systems can also be written in the form
\be
\begin{array}{rl}
a(t) &= \displaystyle\sqrt{\frac{m}{2 \hbar}} \alpha \left(
     \frac{\hbar}{m} \frac{\partial}{\partial x} - i \left(\frac{2 \hbar}{m} y \right) x  \right)\\[3ex]
a^{+}(t)&= \displaystyle\sqrt{\frac{m}{2 \hbar}} \alpha \left(
  -   \frac{\hbar}{m} \frac{\partial}{\partial x}  + i \left(\frac{2 \hbar}{m} y^{\ast}\right) x
     \right).
\end{array}
\label{mex97}
\ee
The corresponding CSs can again be obtained via application of $a(t)$ as eigenstates of this operator or via the displacement operator, leading to 
\be
\vert z\rangle = e^{- \frac{1}{2}  \vert z \vert^{2}} \sum_{n=0}^{\infty} \frac{z^{n} (a^{+})^{n}}{n !}   \vert 0\rangle,
\label{mex98}
\ee
with
\[         
z = \sqrt{\frac{m}{2 \hbar}} \alpha \left[
\left(\frac{2 \hbar}{m} y_{I}\right) \eta + i \left( \dot{\eta} - \left(\frac{2 \hbar}{m} y_{R}\right) \eta \right) \right],
\]
where the CSs $ \vert z\rangle$ are identical to the exact WP solutions $\Psi_{WP}$ of the
NLSEs discussed in Sections 3.3 and 3.4 and related to the canonical WPs of
Sections 3.1 and 3.2 via the non-unitary transformation given in Section 3.5
(for further details, see also \cite{43}).

\section{ Conclusions}

A generalization of CSs and the creation/annihilation operators used to
construct them is necessary if they represent WPs with time-dependent width
corresponding to TD position uncertainties. This situation not only occurs for
HOs with TD frequency $\omega = \omega (t)$, but already for $\omega =
\omega_0 =$ constant if the initial state is not the ground state of the
harmonic oscillator. In these cases, the CS is not a minimum uncertainty CS
(since position/momentum correlations are involved) but still fulfills the
Schr\"odinger--Robertson uncertainty relation. Using a Gaussian WP ansatz,
from the corresponding TDSE the equation of motion for the WP width can be
obtained via a nonlinear Riccati equation for a {\it complex} variable
$\left(\frac{2 \hbar}{m} y (t) \right)$. This equation can be transformed into
an Ermakov equation for a quantity $\alpha (t)$ which is directly proportional
to the WP width. This equation, together with the Newtonian equation that
describes the classical motion of the WP maximum, allows for the definition of
the dynamical invariant $I_L$. Instead of considering the Ermakov equation, we concentrate on the complex Riccati equation and its formal analytical solutions which not only depend on the physical system under consideration, but also qualitatively on the (complex) initial conditions. These solutions have close formal similarities with general superpotentials leading to isospectral potentials in supersymmetric quantum mechanics.

The generalized creation/annihilation operators can be expressed in terms of
the complex Riccati variable using the relation between its imaginary part and
$\omega_0$ occurring in the standard form, i.e., $\left(\frac{2 \hbar}{m} y_I
\right) =  \omega_0$, and replacing $\omega_0$ in $a$ and $a^+$ by the complex
quantity accordingly. The resulting operators are in agreement with those
obtained via factorization of the operator corresponding to $I_L$.

Using the definition of the CS as eigenstate of the annihilation operator, it
is possible to determine the complex eigenvalue $z(t)$ in terms of the
classical quantities $\eta$ and $\dot{\eta}$ and of real and imaginary parts
of $\left(\frac{2 \hbar}{m} y \right)$. The CS in position space obtained in this way is identical to the Gaussian WP that is a solution of the corresponding TDSE.

In addition, the definition of the CS via the displacement operator, expressed
using the generalized $a(t)$ and $a^+(t)$, provides the same Gaussian WP. Finally, the mean value of position and momentum, as well as their uncertainties, are calculated using the generalized CSs and it is proven that the Schr\"odinger--Robertson uncertainty condition is fulfilled.

In the second part, the formalism is extended to include dissipative environmental effects. For this purpose descriptions of the dissipative systems in terms of modified SEs are considered. Specifically, two approaches are discussed based on classical Hamiltonians obtained via non-canonical transformations leading to modified, but linear, TDSEs on a formal canonical level and are compared with two approaches using nonlinear modification of the TDSE. In all cases, Gaussian WP solutions again exist and the relevant complex Riccati equation is the same for both nonlinear approaches. The Riccati equations of the canonical approaches are related via a unitary transformation of the corresponding WPs and can be transformed into the one of the nonlinear approaches taking into account the non-unitary transformation (84) between the canonical and the physical level (where the NLSEs are valid).

In the dissipative case the mean value of position, and hence the WP maximum,
obeys a Newtonian equation with additional friction term linear proportional
to velocity. The position and momentum uncertainties, and their contributions
to the energy of the CS, are also determined. When expressed in terms of
the auxiliary Ermakov variable $\alpha_{NL}(t)$ they have a different form
from the case without dissipation, particularly because the relation between $\left(\frac{2 \hbar}{m} y_R
\right)$ and $\alpha, \dot{\alpha}$ is different. However, expressed in terms
of real and imaginary parts of $\left(\frac{2 \hbar}{m} y \right)$, the form is identical to the one in the conservative case. The same applies to the creation/annihilation operators that can be obtained in the same way as in part one. They also agree with those obtained via factorization of the Ermakov invariant corresponding to the dissipative systems (if the difference between canonical and physical levels is taken into account).

Concluding, one can say that creation and annihilation operators,
corresponding CSs and quantum uncertainties for time-dependent conservative
and dissipative Hamiltonians can all be expressed in the same form if the
complex variable $\left(\frac{2 \hbar}{m} y \right)$ of the Riccati equation
is applied. Due to the nonlinearity, for the same physical system different
CSs with different physical properties can exist according to the respective
solution of the Riccati equation. Furthermore, these solutions can also be
very sensitive to the (complex) initial conditions. In this context, formal
similarities and comparison with supersymmetric quantum mechanics and the CSs
of the nonlinear algebras involved \cite{39,40,41,42} might be useful and will be investigated.

\subsection*{Acknowledgments}
The financial support of CONACyT, projects 101541 and 152574, and IPN project
SIP-SNIC-2011/04 is
acknowledged. This work was carried out when DS visited the Instituto de
Ciencias Nucleares, UNAM and the Departamento de F\'{i}sica, Cinvestav
(Mexico). DS acknowledges these institutions, particularly Octavio Casta\~{n}os and Oscar Rosas-Ortiz, for their kind hospitality.

\end{document}